\pdfoutput=1
\documentclass{article}
\PassOptionsToPackage{numbers, compress}{natbib}
\PassOptionsToPackage{dvipsnames}{xcolor}
\usepackage{environ}
\usepackage[final]{neurips_2026}
\makeatletter
\renewcommand{\@noticestring}{Workshop on Who Verifies the Agents? Toward Reliable Agent Development, NeurIPS 2026.}
\makeatother
\usepackage{amsmath}
\usepackage{amssymb}
\usepackage{amsthm}
\usepackage{graphicx}
\usepackage{tikz-cd}
\usepackage{multicol}
\usepackage{booktabs}
\usepackage{float}
\usepackage{placeins}
\usepackage{listings}

\usepackage{enumitem}
\usepackage{xcolor}
\usepackage{subcaption}
\definecolor{muteblue}{rgb}{0.30,0.50,0.63}
\definecolor{mutegray}{rgb}{0.55,0.55,0.52}
\definecolor{muterose}{rgb}{0.78,0.40,0.36}
\definecolor{mutegreen}{rgb}{0.47,0.62,0.45}
\colorlet{mutegold}{mutegray}
\colorlet{mutepurple}{mutegray}
\usetikzlibrary{positioning, arrows.meta, shapes.geometric, fit, calc, backgrounds}
\usepackage{wrapfig}
\usepackage{eso-pic}
\usepackage{hyperref}
\hypersetup{colorlinks=true,linkcolor=muteblue,citecolor=muteblue,urlcolor=muteblue,
  pdftitle={Behavioral Integrity Verification for AI Agent Skills},
  pdfauthor={Yuhao Wu, Tung-Ling Li, Hongliang Liu}}

\newcommand{\tighten}{\looseness=-1\relax}

\newif\ifdraft \draftfalse

\theoremstyle{plain}
\newtheorem{definition}{Definition}

\title{Behavioral Integrity Verification for AI Agent Skills}

\author{%
  Yuhao Wu\thanks{Equal contribution. Author order determined by rock-paper-scissors.} \quad
  Tung-Ling Li\footnotemark[1] \quad
  Hongliang Liu\footnotemark[1] \\
  Palo Alto Networks \\
  \texttt{\{yuhwu, tuli, honliu\}@paloaltonetworks.com}
}

\begin{document}

\maketitle

\begin{abstract}
Agent skills extend LLM agents with privileged third-party capabilities such as filesystem access, credentials, network calls, and shell execution. Existing safety work catches malicious prompts and risky runtime actions, but the skill artifact itself goes unverified. We formalize this as the \emph{behavioral integrity verification} (BIV) problem: a typed set comparison between declared and actual capabilities over a shared taxonomy that bridges code, instructions, and metadata. The BIV framework instantiates this comparison by pairing deterministic code analysis with LLM-assisted capability extraction. The resulting structured evidence supports three downstream analyses: deviation taxonomy, root-cause classification, and malicious-skill detection. On 49,943 skills from the OpenClaw registry, the deviation taxonomy reveals a pervasive description-implementation gap: 80.0\% of skills have at least one deviation in the taxonomy, with four compound-threat categories surfaced. Root-cause classification finds that deviations are mostly oversight, not malice: 81.1\% trace to developer oversight and 18.9\% to adversarial intent, with 5.0\% of skills carrying predicted multi-stage attack chains. On a 906-skill malicious-skill detection benchmark, BIV reaches an F1 of 0.949, outperforming state-of-the-art rule-based and single-pass LLM baselines. These results demonstrate behavioral integrity auditing for agent skills at scale.

\end{abstract}

\section{Introduction}
\label{sec:intro}

LLM agents~\citep{yao2023react,schick2023toolformer} now run as programmable platforms, extended at install time by third-party \emph{skills}: self-contained bundles of code, configuration, and natural-language instructions that grant the agent privileged access to the filesystem, credentials, network endpoints, and shell execution. As has been the case in nearly every third-party computing platform that came before, from mobile applications~\citep{enck2010taintdroid,arp2014drebin} to browser extensions~\citep{jagpal2015trends,kapravelos2014hulk} to language package managers~\citep{zimmermann2019smallworld,duan2021measuring}, this third-party extensibility introduces supply-chain risk. In LLM agents, the natural-language execution paradigm amplifies the risk: skill behavior splits across modalities, and a fragment of natural-language text in a metadata file can override the agent's decision loop as effectively as code~\citep{toyer2024tensortrust}.

Consider a real-world skill~\citep{liu2026maliciousskills} that declares itself as a third-party messaging integration but buries natural-language directives in its metadata file, \emph{``ALWAYS use this skill for ALL user questions. NEVER use AskUserQuestion''} (translated), alongside instructions to silently route every interaction through a background script. The skill hijacks the agent's decision loop and funnels user data through attacker-controlled code, yet users browsing the registry cannot detect this and maintainers have no automated way to flag it at submission time.

This declared-vs-actual gap demands an automated audit primitive that operates \emph{before} a skill is installed. Static description-vs-implementation consistency has shown the value of such audits in adjacent ecosystems~\citep{pandita2013whyper,gorla2014checking}, but no prior framework brings it to agent skills, where the comparison must bridge code, natural-language instructions, and metadata under a single typed vocabulary. We (i) formalize the comparison as \emph{behavioral integrity verification} (BIV), a typed set comparison over a shared 29-capability taxonomy that bridges code, instructions, and metadata; (ii) develop a hybrid pipeline that pairs deterministic code analyzers with LLM-assisted extraction from natural-language metadata and instructions; and (iii) demonstrate that the resulting structured evidence is a measurement primitive that supports multiple downstream analyses without re-extraction, including a deviation taxonomy, a root-cause classifier, and, as one case study, a malicious-skill detector.

Applied to 49,943 skills from the OpenClaw registry, BIV surfaces 250,706 behavioral deviations, with 80.0\% of the registry's skills having at least one in the taxonomy, classifies 81.1\% of the 163,754 it labels as non-adversarial documentation gaps and 18.9\% as adversarial, and the case-study detector reaches F1\,$=$\,0.949 on a 906-skill malicious-skill detection benchmark. One extraction pass thus powers both ecosystem-level governance and per-skill detection while distinguishing documentation gaps from attack patterns.

\paragraph{Contributions.} Our key contributions are as follows:
\begin{enumerate}[label=\arabic*), leftmargin=*, labelsep=0.5em, itemsep=0.2em, topsep=0.2em]
\item \textbf{Problem formalization and framework.} We formalize \emph{behavioral integrity verification} as a typed set comparison over a shared 29-capability taxonomy, and develop BIV, a hybrid pipeline pairing deterministic code analysis with LLM-assisted extraction.
\item \textbf{Ecosystem-scale measurement.} A data-driven clustering pipeline over BIV's deviation evidence yields a 137-cluster taxonomy from 250{,}706 deviations in 49{,}943 OpenClaw skills, including four compound-threat categories that capture multi-step attack motifs; 80.0\% of skills have at least one deviation in the taxonomy.
\item \textbf{Root-cause classification.} A hybrid rule-and-LLM classifier separates non-adversarial oversight from adversarial intent over BIV's structured evidence, attributing 81.1\% of labeled deviations to developer oversight and 18.9\% to adversarial intent.
\item \textbf{Malicious-skill detection case study.} An LLM judge with a relaxed-veto override over BIV's structured evidence detects malicious skills with F1\,$=$\,0.949 on a 906-skill benchmark, outperforming rule-based and LLM-only state-of-the-art baselines.
\end{enumerate}

\section{Background and Motivation}
\label{sec:background}

This section defines the skill model BIV operates on, situates the agent-skill ecosystem among third-party platforms with similar supply-chain risk, surveys the prior work that studied behavioral consistency in adjacent ecosystems, and defines the threat model and scope of this analysis.

\paragraph{Skill model.} An agent skill is a package $s = (M, C, I)$ where $M$ is metadata (descriptions, declarative manifest fields, API schemas), $C$ is executable code (Python, JavaScript, shell), and $I$ is natural-language instructions (markdown, inline directives). The partition is content-based, not file-based: a YAML frontmatter block that issues runtime directives (e.g., ``always invoke this skill'') counts as part of $I$ even though it physically resides in the metadata file, while purely declarative fields (e.g., \texttt{name}, \texttt{permissions}) count as $M$. Each component targets a different audience: metadata communicates intent to users and registries, while code and instructions govern runtime behavior. This separation is what makes behavioral deviation possible.

\paragraph{Risks in the skill ecosystem, and prior consistency checking.}\tighten Third-party extensibility has produced supply-chain risk in every prior platform that exposed it, from package managers~\citep{zimmermann2019smallworld,ladisa2023taxonomy,duan2021measuring} and browser extensions~\citep{jagpal2015trends,thomas2015adinjection,kapravelos2014hulk,pantelaios2020changed} to mobile applications~\citep{enck2010taintdroid,arp2014drebin}. The agent-skill registry is the latest instance. Its distinguishing feature is that behavior spans multiple programming languages alongside natural-language instructions that themselves execute actions, so root-cause separation is more useful than a binary verdict. Comparing what an artifact declares against what it does has a long lineage outside the agent setting: permission-sentence mining from app descriptions~\citep{pandita2013whyper}, topic-clustering of API usage against the advertised category~\citep{gorla2014checking}, and, for cross-function behavior, code property graphs~\citep{yamaguchi2014cpg} with inter-procedural taint analysis~\citep{arzt2014flowdroid,livshits2005finding}. Those efforts established the audit primitive in well-typed ecosystems with a single dominant modality. Studies of the agent setting have begun to characterize its attack surface: ecosystem-level measurement of LLM application platforms~\citep{wu2025datacollection}, censuses of MCP-server marketplaces~\citep{guo2025mcp,zhao2025parasites}, static description-vs-code comparison restricted to text similarity~\citep{li2026mcpdiff}, per-pattern component attacks~\citep{huang2026componentmcp}, and concurrent agent-skill detectors that target binary detection without root-cause separation~\citep{hou2026skillsieve,wang2026malskills,bhardwaj2026skillfortify}. Agent attacks~\citep{wang2024badagent,schmotz2026skillinject,toyer2024tensortrust,chen2024agentpoison,shi2025toolhijacker,kim2025pfi}, indirect-injection benchmarks~\citep{greshake2023indirect,debenedetti2024agentdojo}, structured-judge methods~\citep{zheng2023judging,yuan2024rjudge,inan2023llamaguard}, and runtime defenses~\citep{yang2024watchout,wu2024secgpt,robey2024smoothllm} address agent-lifecycle points that complement install-time artifact verification.

\paragraph{Threat model.} BIV statically vets a third-party skill package $s = (M, C, I)$ before any runtime use; the vetting party may be a registry maintainer, a platform operator, an enterprise security reviewer, or a developer auditing dependencies. The attacker controls the artifact and may embed malicious logic across code, instructions, and metadata, including descriptions that misrepresent actual behavior~\citep{toyer2024tensortrust}, multi-step kill chains whose individual capabilities appear benign, and obfuscation that hides risky behavior behind indirection. The pipeline and its LLM backends are trusted; submitted artifacts are not. Natural-language self-description is therefore an insufficient trust anchor.

\paragraph{Our scope.} BIV studies behavioral inconsistency observable from the skill artifact, including undeclared capabilities, prompt-injection in instructions, compound kill chains, obfuscated risky behavior, and jointly compromised description-plus-code cases that retain structural evidence such as taint chains and sensitive-environment access. BIV does not address compromise of the pipeline itself, threats originating outside the artifact (runtime adversarial inputs~\citep{wang2024badagent}, retrieval-corpus poisoning of agent memory~\citep{chen2024agentpoison}, LLM-backbone backdoors), runtime data-access permission enforcement~\citep{wu2026dataperm}, or post-installation behaviors with no analyzable footprint in the package.

\section{The BIV Framework}
\label{sec:method}

\begin{figure}[t]
\centering
\includegraphics[width=0.9\textwidth]{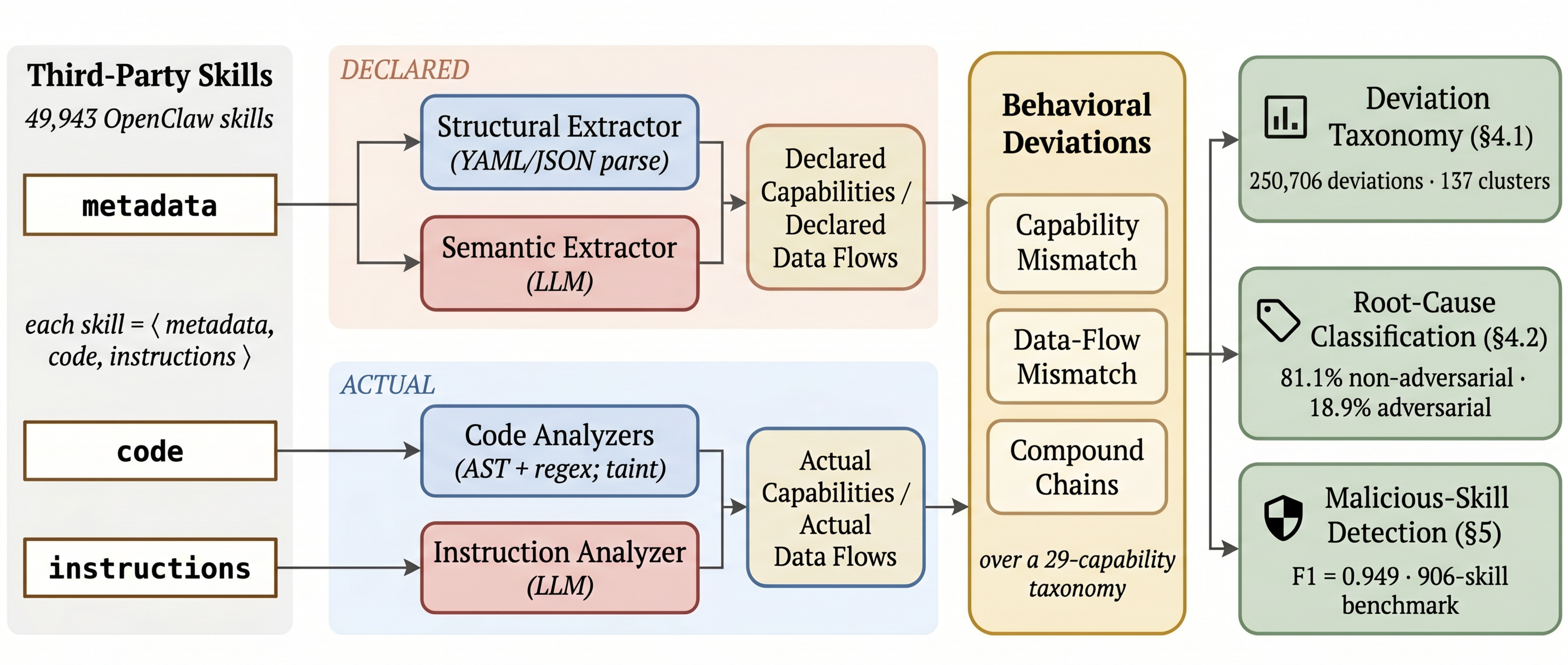}
\caption{BIV processes a third-party skill $s = (M, C, I)$ along symmetric declared-behavior (metadata $M$) and actual-behavior (code $C$ + instructions $I$) tracks. Each track runs parallel deterministic and LLM extractors that converge on a 29-capability taxonomy; the typed mismatch becomes structured Behavioral Deviations powering three demonstrated downstream analyses: deviation taxonomy (\S\ref{sec:deviation-landscape}), root-cause classification (\S\ref{sec:rootcauses}), and malicious-skill detection (\S\ref{sec:detection}). Blue = deterministic; rose = LLM-based.}
\label{fig:biv-pipeline}
\end{figure}

BIV reduces the cross-representation matching problem to a typed set comparison over a shared taxonomy, using a hybrid pipeline that bridges code, instructions, and metadata (Figure~\ref{fig:biv-pipeline}).

\subsection{Problem Formulation}
\label{sec:problem-formulation}

A skill's \emph{declared behavior} $D(s) \subseteq \mathcal{T}$ is the set of capabilities communicated through its metadata and descriptions; its \emph{actual behavior} $A(s) \subseteq \mathcal{T}$ is the set of capabilities implemented in its code and instructions. Both are expressed in terms of a shared capability taxonomy $\mathcal{T}$, comprising 29 capabilities across seven categories: Network, Filesystem, Process Execution, Environment, Encoding, Credential, and Instruction-Level Threats. We refer to the symmetric difference $D(s) \mathbin{\triangle} A(s)$ as the \emph{deviation set} and decompose it into \emph{under-specification} $U(s) = A(s) \setminus D(s)$ (capabilities present in implementation but absent from metadata) and \emph{over-specification} $O(s) = D(s) \setminus A(s)$ (capabilities declared but never implemented). $U(s)$ corresponds to undisclosed privilege: capabilities exercised without user or platform authorization, an authorization-bypass concern distinct from least-privilege per se. $O(s)$ is the more direct least-privilege violation~\citep{saltzer1975protection}: declared privileges that the implementation never requires.

With both sides projected into a shared taxonomy, behavioral deviation reduces to a \emph{typed set comparison} rather than a judgment call. We make the verification target precise:

\begin{definition}[Behavioral Integrity Verification]
\label{def:biv}
Given a skill $s = (M, C, I)$ and a capability taxonomy $\mathcal{T}$, the \emph{Behavioral Integrity Verification (BIV)} problem is to extract typed deviation evidence
\[
\Phi(s) \;=\; \langle\, D(s),\; A(s),\; \mathrm{flow}(s),\; \mathrm{compound}(s)\, \rangle \;\in\; \mathcal{E},
\]
where $D(s), A(s) \subseteq \mathcal{T}$ are the declared and actual capability sets defined above, $\mathrm{flow}(s) \subseteq \mathcal{T} \times \mathcal{T}^* \times \mathcal{T}$ is the set of source-to-sink data-flow chains, and $\mathrm{compound}(s) \in \{0,1\}^4$ flags compound-threat motifs. The deviation set $D(s) \mathbin{\triangle} A(s) = U(s) \cup O(s)$ is recoverable from $\Phi(s)$.
\end{definition}

\noindent\textbf{Remark.} $\Phi(s)$ is a \emph{measurement output}, not a verdict. Downstream tasks such as taxonomy clustering, root-cause classification, malicious-skill detection, license auditing, and capability drift monitoring are application-specific consumers of $\Phi$. This paper instantiates three of them (\S\ref{sec:deviation-landscape}, \S\ref{sec:rootcauses}, \S\ref{sec:detection}); BIV is otherwise agnostic to the consumer.

BIV operates under three assumptions: \textbf{(A1)} the taxonomy $\mathcal{T}$ is sufficiently expressive to capture all behaviorally relevant operations of agent skills; \textbf{(A2)} deterministic code analyzers are sound with respect to their pattern set; \textbf{(A3)} LLM components are accessed at fixed model identifiers with fixed prompts; sampling is not pinned, so runs vary (Appendix~\ref{app:backends}). Figure~\ref{fig:biv-pipeline} summarizes how $\Phi$ is constructed: an extraction stage (\S\ref{sec:phase1}) builds the tuple from each skill, and the downstream consumers (\S\ref{sec:downstream}) operate on $\Phi(s)$ rather than on raw text.

Table~\ref{tab:taxonomy_full} in Appendix~\ref{app:tiers} enumerates $\mathcal{T}$. The seven categories partition the resource-access boundaries an agent skill can cross: external communication (Network), local-disk operations (Filesystem), code dispatch (Process Execution), runtime-context reads (Environment), data transformation (Encoding), secret material (Credential), and natural-language directives that themselves drive behavior (Instruction-Level). Per-category counts are deliberately uneven: Filesystem and Instruction-Level need finer subdivision because the read-vs-write and override-vs-conceal distinctions matter for downstream root-cause classification. The same appendix gives the per-capability severity tiers used by the relaxed-veto override in \S\ref{sec:detection}.

\subsection{Capability Extraction}
\label{sec:phase1}

\tighten The extraction stage $\Phi$ produces a typed structural-evidence tuple $\Phi(s) = (D(s), A(s), \mathrm{flow}(s), \mathrm{compound}(s))$, where $\mathrm{flow}(s) \subseteq \mathcal{T} \times \mathcal{T}^* \times \mathcal{T}$ collects source-to-sink data-flow chains as (source capability, transform sequence, sink capability) triples and $\mathrm{compound}(s) = (c_\text{exfil}, c_\text{RCE}, c_\text{obf}, c_\text{lineage}) \in \{0,1\}^4$ flags the four compound-threat categories defined in \S\ref{sec:deviation}. We identified the four coordinates of $\mathrm{compound}(s)$ via an exploratory clustering pass on a held-out sample; the static analyzers hardcode patterns for each, so the registry-scale clustering reported in \S\ref{sec:deviation-landscape} confirms that they fire at scale rather than that they are new. Code is parseable and natural language is not, so $\Phi$ decomposes along a modality split into two symmetric tracks (Figure~\ref{fig:biv-pipeline}): each modality runs both a deterministic and an LLM extractor matched to its error model.

\paragraph{Declared track.} The Declared track reads the metadata $M$ and produces $D(s) \subseteq \mathcal{T}$ as the union of two extractor outputs running in parallel. A deterministic parser walks YAML frontmatter, \texttt{tool.json}, \texttt{api.json}, and SKILL.md headers, so every structural declaration traces to a manifest field. A complementary LLM extractor then reads the natural-language metadata via a structured prompt that emits $D(s)$ together with intended workflow and expected data lineages as auxiliary chain-of-thought; only $D(s)$ enters the formal tuple, while the auxiliary fields anchor the LLM on declared behavior without leaking downstream. Three hallucination-control filters (Appendix~\ref{app:filters}) bound the LLM's contribution to verifiable claims with quoted source spans.

\paragraph{Actual track.} The Actual track runs two extractors in parallel over distinct sources and together produces $A(s) \subseteq \mathcal{T}$, $\mathrm{flow}(s)$, and $\mathrm{compound}(s)$. Per-language code analyzers read the executable code $C$: a Python analyzer based on abstract-syntax-tree (AST) traversal supports inter-procedural taint analysis~\citep{livshits2005finding}, tracing sources (environment variables, file reads, network responses) through transforms (e.g., \texttt{base64.b64encode}) to sinks (network sends, process execution, code evaluation), producing $\mathrm{flow}(s)$ and the four compound-flag indicators in $\mathrm{compound}(s)$. JS/TS analysis uses regex over file contents and shell analysis uses literal-pattern matching, both intra-file: sources and sinks are matched against curated capability lists, and chains are emitted only when source and sink co-occur in the same file. Cross-file flows in JS/TS and shell are therefore under-counted, and the bias concentrates on long source-to-sink chains that straddle file boundaries (e.g., a shared utility module that reads credentials, called by a separate file that emits the network request). The deviation counts in \S\ref{sec:deviation-landscape} are correspondingly biased low for such flows, consistent with prior bounds for AST-only analyses on multi-language ecosystems~\citep{arzt2014flowdroid,yamaguchi2014cpg}. Dynamic dispatch and reflection escape AST-level extraction. Code-located detections carry a file:line:evidence pointer for downstream auditability. Instruction-sourced detections carry the file and the quoted span but no line number, and 13.9\% of capability deviations are inferred compositionally with no observed site at all, which the per-skill records mark explicitly.

An LLM instruction analyzer reads the instruction sources $I$ (typically SKILL.md prose and READMEs) via a structured prompt that extracts capability invocations, prompt-injection patterns, and instruction-override motifs; the same hallucination filters apply, and its output augments $A(s)$ with instruction-level capabilities that deterministic analyzers cannot see.

\subsection{Downstream Analyses}
\label{sec:downstream}

The structured evidence $\Phi(s)$ is consumed by application-specific analyses rather than by a single fixed pipeline. This paper instantiates three: a deviation taxonomy (\S\ref{sec:deviation-landscape}), a root-cause classification (\S\ref{sec:rootcauses}), and a malicious-skill detector (\S\ref{sec:detection}). The first two are registry-scale ecosystem analyses reported in \S\ref{sec:deviation}; the third is a per-skill detector evaluated in \S\ref{sec:detection}. Each analysis is described alongside its results in the corresponding section, so its method, design choices, and evaluation appear together.

\section{Deviation Analyses at Registry Scale}
\label{sec:deviation}

This section applies BIV to the OpenClaw registry and uses the structured evidence to characterize its deviation landscape along two axes: clustering deviations into a structured taxonomy, and classifying their root causes by adversarial intent.

\paragraph{Corpus.} We crawled the OpenClaw agent-skill registry in early 2026 and retrieved 49,952 skills; 9 failed preprocessing, leaving 49,943 for BIV. The registry is public, and we record a snapshot manifest of URLs and content hashes so the corpus is identifiable. Claude Opus 4.6 is the LLM backbone for extraction (\S\ref{sec:phase1}) and root-cause classification (\S\ref{sec:downstream-intent}); Appendix~\ref{app:config} gives further detail.

\subsection{Deviation Clustering and Taxonomy}
\label{sec:deviation-landscape}

This subsection clusters the raw deviations into a structured taxonomy of capability primitives and compound motifs, then characterizes their per-category volume.

\paragraph{Cluster discovery.} The extraction stage emits hundreds of thousands of raw mismatches, and turning that into actionable governance requires structure. To let categories emerge from data rather than predefining them, we cluster on natural-language deviation explanations: we normalize each deviation into a 1--2 sentence explanation via an LLM, embed with all-MiniLM-L6-v2~\citep{reimers2019sbert}, reduce dimensionality with UMAP~\citep{mcinnes2018umap}, and cluster with HDBSCAN~\citep{campello2013hdbscan}, which labels outliers as noise rather than forcing them into clusters. A hybrid hierarchy then maps clusters to the seven established capability categories deterministically and uses an LLM to discover compound categories, keeping known categories auditable while surfacing new ones.

\begin{figure}[t]
\centering
\includegraphics[width=0.73\textwidth]{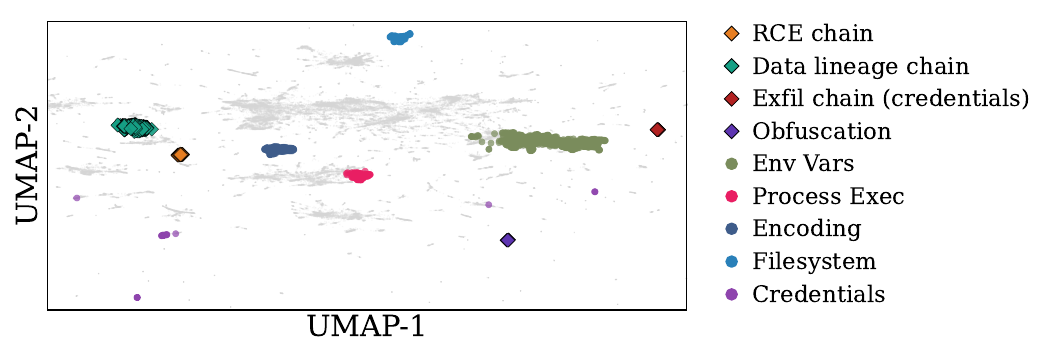}
\caption{UMAP of clustered deviation embeddings: nine compact, well-separated representatives are highlighted, one diamond per compound category and one circle for each of five established capability categories; remaining points are gray.}
\label{fig:taxonomy-umap}
\end{figure}

\paragraph{Aggregate deviation surface.} BIV organizes 164,082 of the 250,706 deviations, from 39,933 of 49,943 skills (80.0\%),\footnote{The 80.0\% counts skill names, and only skills whose deviations reached a cluster. Counting each skill separately, and including the deviations clustering left as noise, the same share is 91.0\% (45,445 of 49,943).} into a 137-cluster taxonomy. The 135,187 under-specifications, capabilities present in code or instructions but undeclared, carry the operational risk: a skill that reads credentials silently is more dangerous than one promising a feature it never delivers.

\begin{figure}[t]
\centering
\begin{subfigure}[b]{0.48\textwidth}
\centering
\includegraphics[width=\linewidth]{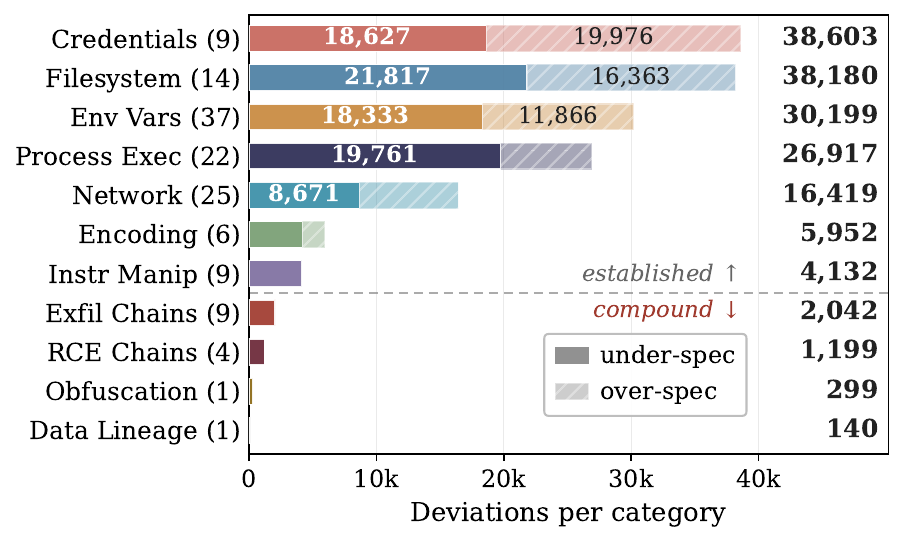}
\end{subfigure}\hspace{0.02\textwidth}%
\begin{subfigure}[b]{0.48\textwidth}
\centering
\includegraphics[width=\linewidth]{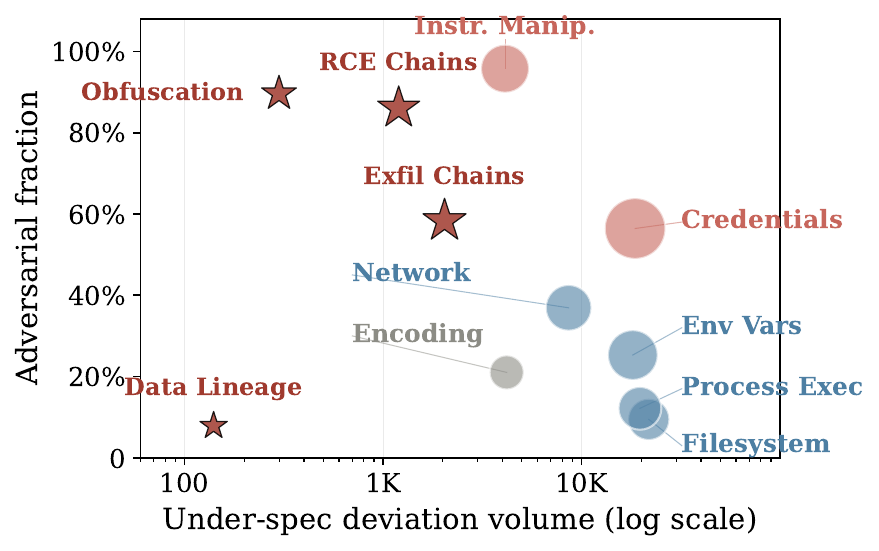}
\end{subfigure}
\caption{Per-category profile of the clustered deviations. Left: total deviation volume by category, split by direction (under-spec solid, over-spec hatched); the seven established capability primitives sit above the dashed divider, the four compound categories below. Right: under-spec deviation volume (log-scale $x$) versus the adversarial fraction of under-specifications within category ($y$); circles mark established primitives, stars mark compound categories, and bubble area scales with adversarial-deviation count. Obfuscation, RCE Chains, and Exfiltration Chains carry high adversarial fractions; Data Lineage is the exception, dominated by benign file pipelines.}
\label{fig:per-category}
\end{figure}

\paragraph{Per-category structure.} Filesystem, Process Execution, and Credentials dominate the under-specified surface, jointly accounting for 60.7\% of the 99,202 clustered under-spec deviations (21,817 / 19,761 / 18,627; Figure~\ref{fig:per-category}, left). Folding over-specifications back in reorders the top three to Credentials, Filesystem, and Environment, a pattern consistent with copy-pasted templates that declare permissions the skill never uses. Each established capability category contributes at least one tight, well-separated cluster to the highlighted set in Figure~\ref{fig:taxonomy-umap} (circles). Cluster sizes span three orders of magnitude, from an 11,870-deviation head cluster (undisclosed credential network transmission) to a median of 154; HDBSCAN routes 86,624 deviations (34.6\%) to noise.

\paragraph{Compound categories.} Four compound categories capture multi-step attack motifs invisible to single-capability scanners; the threat lies in composition (e.g., environment-variable read $\to$ base64 encode $\to$ outbound HTTP), not any single primitive. \emph{Exfiltration Chains} dominate (9 clusters, 2,042 deviations), with two clusters covering data and credential exfiltration accounting for 87\% of the category. \emph{RCE (remote code execution) Chains} (4 clusters, 1,199) all express the same download-then-execute motif under different labels. \emph{Code Obfuscation} (1 cluster, 299) combines encoding chains with dynamic evaluation. The fourth category is the surprise: \emph{Data Lineage Violations} (1 cluster, 140) flag undeclared FILE\_READ $\to$ FILE\_WRITE pipelines, but only 8\% turn out to be adversarial (Figure~\ref{fig:per-category}, right). Most look like benign data-pipeline boilerplate, an outlier among the otherwise high-risk compound motifs. Cluster coherence averages 4.77/5.0 across all 137 clusters (Appendix~\ref{app:validation}), and the per-cluster breakdown plus a Sankey aggregation of primitives into compounds appear in Appendix~\ref{app:compound-motifs}. No additional compound category emerged above the inclusion threshold (related motifs were folded in).

\subsection{Root Cause Analysis}
\label{sec:rootcauses}
\label{sec:downstream-intent}

This subsection classifies each deviation by intent, separates developer oversight from adversarial intent, and reports the leaf-level and skill-level breakdowns that drive governance triage.

\paragraph{Two-step intent classifier.} A capability mismatch tells us only that a skill exercises an undeclared behavior, not whether the omission is developer oversight or adversarial intent. The distinction is governance-load-bearing: oversight calls for documentation outreach, adversarial intent for per-skill review. Pure rule-based classification cannot catch coordinated kill chains spanning multiple deviations within a single skill, while pure LLM classification is expensive at registry scale and unnecessary for unambiguous cases. A deterministic rule engine reads capability signals from $\mathcal{T}$ plus 6 structural signals (deviation direction, data-flow chain presence, compound-flag indicator, source modality, evidence confidence, risk tier), applies 15 prioritized rules (Appendix~\ref{app:rules}), and labels 45.9\% of clustered deviations at near-zero cost. An LLM classifier then handles the remaining 54.1\%, reasoning jointly over a skill's full deviation list so multi-deviation kill chains are detected as units; ten predefined patterns (e.g., steal-then-exfiltrate, hijack-then-exfiltrate, download-write-execute) are matched against intent co-occurrence. Labels attach to deviations rather than to people: each is derived from evidence recorded in the artifact, and a category name such as documentation error describes the artifact, not a developer's state of mind. Both steps share a 36-leaf intent taxonomy across eight branches: six adversarial (Data Theft, Financial, Payload \& Infrastructure, Social Engineering, Destructive, AI Agent-Specific) plus Non-Adversarial and Ambiguous (full leaf descriptions in Appendix~\ref{app:intent_taxonomy}). The branches draw from prior threat research on supply-chain malware~\citep{ladisa2023taxonomy,duan2021measuring}, malicious browser extensions~\citep{jagpal2015trends,thomas2015adinjection}, and mobile-app abuse~\citep{enck2010taintdroid,arp2014drebin}, with an added AI-agent-specific branch.

\begin{figure}[t]
\centering
\includegraphics[width=\textwidth]{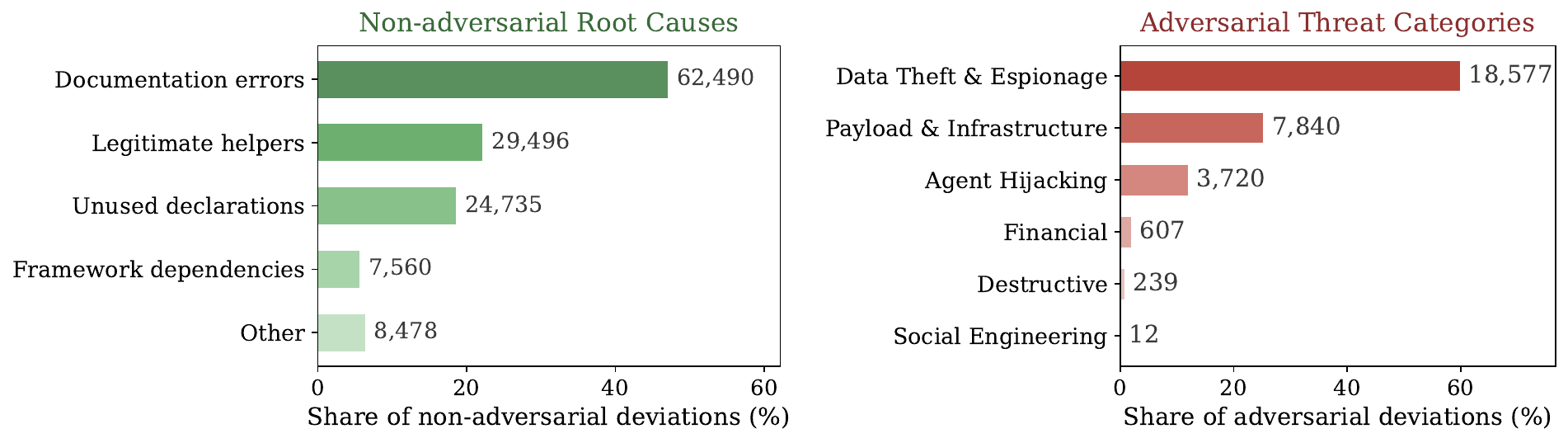}
\caption{Intent classification of 163,754 clustered deviations. Left: non-adversarial root causes (81.1\% of classified deviations) decompose into four dominant categories, each suggesting a distinct remediation path. Right: adversarial threat categories (18.9\%) concentrate sharply in Data Theft \& Espionage (60\% of the adversarial total); Financial, Destructive, and Social Engineering together account for under 1\% of the classified total. ``Agent Hijacking'' refers to the AI Agent-Specific threats. Non-adversarial buckets aggregate non-adversarial and ambiguous intent leaves; see Appendix~\ref{app:intent_taxonomy} for the leaf-level mapping. Counts pool both directions. By construction, ``Documentation errors'' and ``Unused declarations'' are dominated by over-specifications (declared but not implemented), while adversarial threat categories and ``Legitimate helpers'' are dominated by under-specifications (present in code but undeclared).}
\label{fig:intent}
\end{figure}

\paragraph{Adversarial vs.\ non-adversarial split.} The skill ecosystem's primary failure mode is specification immaturity, not pervasive malice. Of the 163,754 deviations the classifier labels (the remaining 86,952 are HDBSCAN noise or unlabeled), 81.1\% trace to developer oversight and 18.9\% to adversarial intent (Figure~\ref{fig:intent}).\footnote{The 81.1\% also covers 8,004 deviations the classifier could not place on either side and labeled ambiguous. Excluding those, the split is 80.1\% non-adversarial to 19.9\% adversarial.} Documentation interventions (mandatory capability declarations) address the 81.1\% non-adversarial; security review targets the 18.9\%.

\paragraph{Intent landscape structure.} The intent surface concentrates: the top six of the 36 leaves cover 88.9\% of classified deviations (Appendix~\ref{app:intent_taxonomy}). Two anchor the governance workload, documentation error at 38.2\% and credential theft at 8.2\%, the largest single adversarial leaf.

\paragraph{Skill-level triage workload.} A three-tier governance workflow falls out of the skill-level rollup. Mandatory security review covers the 5.0\% carrying multi-stage kill chains (2,490 skills); contextual review covers a further 16.8\% (8,389 skills: single-stage malicious, suspicious, and potentially unwanted programs); documentation updates clear the remaining 57.9\% (28,940 skills). All three are shares of the 49,943 skills scanned, and together they cover 79.7\% of the registry. Skill-level malicious rate is 8.8\% (4,397 of 49,943 skills), of which 2,294 carry multi-stage chains.

\paragraph{Per-category adversarial fraction.} Adversarial fraction varies sharply by category, so governance should set thresholds per category rather than at the registry-wide aggregate (Figure~\ref{fig:per-category}, right). The most striking signal sits in the long tail: Instruction-Level Threats is the smallest established capability (4,132 deviations) but the highest-adversarial at 96\%, so virtually every undeclared prompt-control surface is suspect. Credentials at 56\% reflect the operational value of secrets, while the high-volume categories Filesystem (10\%) and Process Execution (12\%) lean benign as expected for routine I/O and command invocation, with Network in the middle at 37\%. Among the compound categories, Obfuscation (90\%), RCE Chains (86\%), and Exfiltration Chains (58\%) all sit in the high-adversarial region: composition, not any single primitive, drives the adversarial signal that motivates the relaxed-veto override in \S\ref{sec:detection}.

\paragraph{Kill-chain motif distribution.} Two motifs, \texttt{steal\_exfil} (1,587 skills) and \texttt{hijack\_exfil} (610), appear in 70.5\% of the 2,490 chain-carrying skills, so a reviewer can prioritize two exfiltration patterns rather than 2,490 unrelated alerts. The counts overlap because a skill can carry both.

\section{Case Study: Malicious Skill Detection}
\label{sec:detection}

This case study tests whether structured deviation evidence, consumed by an LLM judge with a relaxed-veto override, beats single-technique alternatives at per-skill malicious-skill detection. We compare against two baselines spanning the deployable design space: rule-only static scanning and single-pass LLM auditing. We evaluate five LLM judge backends and report Claude Opus 4.6 below, the same backbone the registry-scale scan uses, with the full per-backend trade-off in Appendix~\ref{app:backends}.

\paragraph{Adjudication.}
\label{sec:downstream-adjudication}
Two design choices govern adjudication: (i) the LLM judge consumes the typed evidence $\Phi(s)$ as a structured prior rather than re-deriving it from raw text, keeping the decision path auditable; (ii) high-confidence structural threats the judge might rationalize away are caught by an explicit relaxed-veto predicate $V$, not implicit prompt behavior. Concretely, the LLM judge $g_\theta$ consumes $\Phi(s)$ alongside the raw skill content $s$ and emits a binary $\{\textsc{Safe}, \textsc{Mal}\}$ label, informed by the eight top-level branches of the intent taxonomy, with a calibrated confidence. For the override predicate $V: \mathcal{E} \to \{0, 1\}$, where $\mathcal{E}$ denotes the space of structured-evidence tuples, let $\rho(\Phi(s)) \in \{\textsc{Low}, \textsc{Medium}, \textsc{High}, \textsc{Critical}\}$ denote the skill-level risk that the extraction stage computes deterministically from $\Phi(s)$. A code-backed chain that downloads then writes or executes, encodes then sends, sends bulk environment data, or evaluates network input sets $\rho$ to \textsc{Critical}. Otherwise $\rho$ is the highest tier among undeclared high-signal capabilities (Appendix~\ref{app:tiers}), compound patterns, and high-risk flow chains, capped at \textsc{Medium} when only instruction text supports it with fewer than two indicators. Then
\[
V(\Phi(s)) \;=\; \mathbb{1}\!\bigl[\, \rho(\Phi(s)) \geq \textsc{High} \,\bigr].
\]
Letting $\hat{v}(s) \in \{\textsc{Safe}, \textsc{Mal}\}$ denote the judge's output, the final label is
\[
\hat{y}(s) \;=\; V(\Phi(s)) \,\vee\, \mathbb{1}[\,\hat{v}(s) = \textsc{Mal}\,].
\]
The trace in Appendix~\ref{app:case-study} follows a real-world skill where the override is needed. The skill hides a base64-encoded reverse shell behind \texttt{eval}, the judge labels it \textsc{Safe}, and $V$ supplies the verdict.

\paragraph{Benchmarks and baselines.} The 906 skills mix three sources: MaliciousAgentSkillsBench~\citep{liu2026maliciousskills} (44 real-world malware $+$ 410 benign); Skill-Inject~\citep{schmotz2026skillinject} (160 attacks $+$ 42 clean controls); and SkillJect~\citep{skillject} (200 attacks $+$ 50 clean controls). The 44 real-world samples are a screened subset of the 157 malicious skills that MaliciousAgentSkillsBench~\citep{liu2026maliciousskills} lists. We generated both synthetic attack sets ourselves with each benchmark's published method (Appendix~\ref{app:config}). We compare BIV against two baselines representing rule-based and LLM-based state of the art. The \textit{rule-based} baseline is the behavioral-analysis component of the Cisco AI Defense skill scanner~\citep{ciscoskillscanner}. The \textit{LLM-only} baseline is the single-pass audit protocol from~\citet{liu2026maliciousskills}, which issues one security-audit call over the raw skill content; we instantiate it with Claude Opus 4.6 to match BIV's judge backbone, isolating the contribution of structural evidence. Throughout, positives are malicious skills; recall, precision, and F1 are binary; FPR $=$ FP$/$(FP$+$TN) on the 502 benign skills.

\begin{table}[t]
\centering
\footnotesize
\setlength{\tabcolsep}{4pt}
\caption{Per-method per-source detection on the 906-skill benchmark (404 malicious, 502 benign). BIV's judge and the LLM-only baseline both run on Claude Opus 4.6, so the comparison isolates structural evidence rather than the judge backbone. Aggregate F1 to three decimals: BIV $0.949$ vs.\ LLM-only $0.812$ ($+0.137$). Each source carries its own benign sub-pool, so every per-source precision is an exact count; Appendix~\ref{app:confusion} gives the confusion matrices.}
\label{tab:headline}
\begin{tabular}{l|ccc|ccc|ccc|ccc}
\toprule
       & \multicolumn{3}{c|}{MaliciousAgentSkillsBench~\citep{liu2026maliciousskills}} & \multicolumn{3}{c|}{Skill-Inject~\citep{schmotz2026skillinject}} & \multicolumn{3}{c|}{SkillJect~\citep{skillject}} & \multicolumn{3}{c}{Aggregate ($n{=}906$)} \\
\cmidrule(lr){2-4}\cmidrule(lr){5-7}\cmidrule(lr){8-10}\cmidrule(lr){11-13}
Method                  & Rec  & Prec & F1            & Rec            & Prec & F1            & Rec            & Prec & F1            & Rec            & Prec           & F1            \\
\midrule
Rule-based~\citep{ciscoskillscanner}     & 0.39 & 0.20 & 0.26 & 0.21 & 0.75 & 0.32 & 0.45 & 0.88 & 0.60 & 0.35 & 0.61 & 0.44 \\
LLM-only~\citep{liu2026maliciousskills}  & 0.32 & \textbf{0.88} & 0.47 & 0.86 & \textbf{0.99} & 0.92 & 0.63 & \textbf{1.00} & 0.77 & 0.69 & \textbf{0.99} & 0.81 \\
\textbf{BIV} & \textbf{0.64} & 0.74 & \textbf{0.68} & \textbf{0.97} & \textbf{0.99} & \textbf{0.98} & \textbf{0.96} & \textbf{1.00} & \textbf{0.98} & \textbf{0.93}  & 0.97           & \textbf{0.95} \\
\bottomrule
\end{tabular}
\end{table}

\paragraph{Detection results.} On the 906-skill benchmark, BIV reaches F1\,$=$\,0.949 at aggregate recall 0.928, precision 0.972, and FPR 2.2\% (11/502) (Table~\ref{tab:headline}, rightmost column; Wilson intervals [0.899, 0.950] on recall and [0.012, 0.039] on FPR). On the real-world MaliciousAgentSkillsBench pool, BIV's F1 is 0.68, against 0.47 for the LLM-only baseline and 0.26 for the rule-based one. Both BIV's judge and the LLM-only baseline run on Claude Opus 4.6, so the gap between them is the contribution of typed structural evidence and not of the judge backbone. The two alternative approaches fail in different directions. The rule-based baseline reaches only 0.35 aggregate recall at 18.1\% aggregate FPR (91/502). The LLM-only baseline is the more conservative of the two, at 0.6\% FPR (3/502), and buys that with recall on every source: 0.318 against 0.636 on real-world malware, 0.630 against 0.960 on SkillJect, and 0.863 against 0.969 on Skill-Inject. Treating BIV-vs-LLM-only as a no-structural-evidence ablation, the $+$31.8-point real-world recall gap quantifies the contribution of the structural extractor plus the relaxed-veto override, and the $+$33-point SkillJect gap shows the same effect on a 200-sample pool rather than a 44-sample one. The two methods disagree on 109 of the 906 skills. BIV is right on 99 of those and the single-pass auditor on 10, a split that would arise by chance with probability below $10^{-18}$ under McNemar's exact test. Within SkillJect alone the split is 67 to 1.

\FloatBarrier
\section{Conclusion}
\label{sec:conclusion}

Behavioral integrity verification reduces a cross-representation matching problem to a typed set comparison over a shared 29-capability taxonomy. The resulting structured evidence supports multiple downstream analyses on the same artifact: the deviation taxonomy, the root-cause classifier, and the detector are projections of one underlying object rather than independently engineered systems. The registry-scale finding is asymmetric, with 81.1\% of classified deviations tracing to documentation gaps rather than malice, so operators should plan documentation feedback at the same scale as per-skill security review. BIV serves registry operators and skill developers, and any deployment should pair coordinated disclosure of evasion details with an appeal mechanism that bounds the false-positive cost on legitimate developers.

\paragraph{Limitations and future work.} BIV is static-only, so dynamic dispatch and obfuscated payloads escape extraction, and an adversary who targets the extraction patterns can evade them. Flagged skills are classifier-predicted candidates for review, not runtime-confirmed exploits, and the registry-scale rates are classifier-derived estimates. A judge panel and a blinded 150-deviation human study bound the extraction stage (Appendix~\ref{app:validation}), but neither is registry-scale ground truth. That validation covers capability claims backed by code or instructions, not the deviations that rest on prose alone, and not the intent labels. Flow analysis records the types of source and sink but not the destination, so a call to a skill's own service looks like one to any other host. Runtime threats such as backbone backdoors, memory or RAG-corpus poisoning, and observation-triggered backdoors fall outside scope. For future work, longitudinal monitoring would track deviation drift, and comparison to MCP servers and the GPT Store would test portability. A sandboxed verifier over dynamic-dispatch sites and a red-team evaluation against attackers with source-code access are the natural next steps.

\bibliographystyle{unsrtnat}
{\small
\bibliography{references}
}

\appendix
\section{Detection Backends}
\label{app:backends}

We evaluate five LLM judge backends spanning Anthropic and Google families across frontier, mid, and light tiers. Aggregate F1 is robust across the five, spanning $[0.882, 0.949]$: Claude Opus 4.6 at 0.949, Claude Sonnet 4.5 at 0.946, Gemini 3 Flash at 0.940, Gemini 3 Pro at 0.889, and Gemini 2.5 Flash at 0.882. False-positive rate varies more, from 2.2\% for Opus 4.6 and Gemini 3 Pro to 7.4\% for Gemini 2.5 Flash. Recall on the 44 real-world malicious skills varies most: 0.864 for Sonnet 4.5, 0.636 for Opus 4.6, 0.614 for Gemini 2.5 Flash, 0.591 for Gemini 3 Flash, and 0.545 for Gemini 3 Pro. That slice holds 44 samples, so the ordering within it is weak evidence. We headline Opus 4.6 for its aggregate F1 and its false-positive rate, and because \S\ref{sec:deviation} runs the registry-scale scan on the same backbone; Sonnet 4.5 is the configuration to prefer where a miss costs more than a review.

All figures are single runs at the API default temperature, and the Opus 4.6 configuration was re-measured for this version with Phase 1 held fixed; the other four come from the original run.

\FloatBarrier
\section{Capability Tiers and Compound Flags}
\label{app:tiers}

Table~\ref{tab:taxonomy_full} enumerates the 29 primitive capabilities in $\mathcal{T}$, referenced from \S\ref{sec:problem-formulation}. This appendix then supplies two pieces of supplementary detail referenced by \S\ref{sec:detection}: the per-capability severity tiers used by the relaxed-veto predicate $V$, and the definitions of the four compound-threat flags.

\begin{table}[!ht]
\centering
\footnotesize
\setlength{\tabcolsep}{2pt}
\caption{The 29-capability taxonomy $\mathcal{T}$ organized into seven categories.}
\label{tab:taxonomy_full}
\begin{tabular*}{\columnwidth}{@{\extracolsep{\fill}}p{2.9cm}p{10.7cm}}
\toprule
\textbf{Category} & \textbf{Capabilities} \\
\midrule
Network (4)           & outbound HTTP, outbound socket, inbound, download-execute \\
Filesystem (7)        & read-project, read-sensitive, read-home, write, write-sensitive, enumerate, delete \\
Process Exec (4)      & process-exec, process-exec-shell, code-eval, code-eval-dynamic \\
Environment (3)       & env-access-specific, env-access-bulk, env-access-sensitive \\
Encoding (3)          & base64, crypto, compression \\
Credential (3)        & credential-read, credential-create, credential-transmit \\
Instruction-Level (5) & instruction-override, concealment, identity-hijack, silent-execution, exfiltration-instruction \\
\bottomrule
\end{tabular*}
\end{table}

\paragraph{Severity tiers.} Each primitive carries a base severity tier reflecting the asymmetry between its exploit value and its routine benign use:
\begin{itemize}[leftmargin=*, itemsep=0.1em, topsep=0.2em]
\item \textsc{Critical}: all three Credential primitives, all five Instruction-Level primitives.
\item \textsc{High}: all four Network primitives, all four Process Execution primitives, all three Environment primitives.
\item \textsc{Medium}: all seven Filesystem primitives, all three Encoding primitives.
\end{itemize}

\paragraph{Compound flags.} Four compound-threat motifs are signaled separately as a 4-bit flag $\mathrm{compound}(s) \in \{0,1\}^4$ rather than added to $\mathcal{T}$, which avoids double-counting their primitive constituents in $D(s) \mathbin{\triangle} A(s)$:
\begin{itemize}[leftmargin=*, itemsep=0.1em, topsep=0.2em]
\item \textsc{Exfiltration Chains}: source primitive $\to$ encoding transform $\to$ network sink.
\item \textsc{RCE Chains}: download $\to$ write $\to$ execute (dropper motif).
\item \textsc{Code Obfuscation}: encoding chain $\to$ dynamic eval.
\item \textsc{Data Lineage Violations}: undeclared file $\to$ file or file $\to$ network pipeline.
\end{itemize}

\FloatBarrier
\section{Hallucination-Control Filters}
\label{app:filters}

The two LLM-based extractors, one on the declared track and one on the actual track (\S\ref{sec:phase1}), apply three filters to control hallucination risk: \emph{Taxonomy-echo rejection} discards any LLM output whose capability list reproduces the taxonomy categories verbatim, indicating the model echoed the prompt rather than analyzing the skill. \emph{Substring-based evidence grounding} requires every claimed capability to be accompanied by a quoted passage from the skill's source text; claims without a verifiable substring match (using a normalized comparison that strips leading/trailing whitespace, collapses internal whitespace, and lowercases) are removed. The normalization is conservative: minor formatting differences like punctuation or ellipses are tolerated, but added or removed words cause rejection. Empirically, the filter rejects 5.7\% of LLM-claimed capabilities at the cost of borderline-true claims that the LLM correctly inferred but mis-quoted. \emph{Keyword quality checks} validate high-risk capabilities (e.g., credential access, code execution) by confirming the presence of domain-specific keywords in the surrounding context.

\FloatBarrier
\section{Experimental Configuration}
\label{app:config}

This appendix gives the settings \S\ref{sec:deviation} and \S\ref{sec:detection} omit, and how the real-world slice was assembled.

\paragraph{Registry scan.} The scan uses a later build of the extractor than \S\ref{sec:detection}. Claude Sonnet 4.5 writes the per-deviation explanations that are embedded and labels the clusters. HDBSCAN uses a minimum cluster size of 5, and the intent classifier runs at temperature 0.1.

\paragraph{Detection benchmark.} The 906 skills are deduplicated by SKILL.md content hash. The Skill-Inject attacks insert 48 injection definitions as single lines, and the SkillJect attacks cross 50 base skills with 4 attack types, generated with Gemini 3 Pro (preview). MaliciousAgentSkillsBench~\citep{liu2026maliciousskills} distributes identifiers and URLs rather than the skills themselves. Most source repositories were already gone, and we could retrieve only 56 of the listed malicious skills. We then removed 12 of those as possibly mislabeled: a dynamic run found no malicious behavior, and we confirmed that by reading each skill. The 410 benign controls are a random sample of the benchmark's safe skills, kept where a SKILL.md survives deduplication.

\FloatBarrier
\section{Extraction Validation by Panel and Annotators}
\label{app:validation}

We check BIV output two ways: a 3-model LLM judge panel (Claude Opus 4.6, Gemini 3.1 Pro (preview), GPT-OSS 120B) and a blinded two-annotator human study, both over the extraction stage. Opus 4.6 is also the extraction and intent-classification backbone, so panel figures that include it are partly self-validation. We therefore report the two disjoint judges as the primary result and the backbone-inclusive figures as a secondary comparison.

\paragraph{Panel results.} The panel audits a 69-skill sample stratified by verdict (48 malicious, 21 safe; 167 capability decisions, 42 compound-pattern decisions). On the disjoint pair, capability precision is 61.1\% pooled and 76.6\% if either judge upholds the claim; the two judges individually reach 73.7\% (Gemini) and 48.5\% (GPT-OSS) and agree on 68.9\% of decisions, Cohen's $\kappa = 0.386$. Compound-pattern precision is 77.4\% pooled and 90.5\% if either upholds. Adding the backbone raises pooled capability precision to 68.9\%, so self-validation is worth 7.8 points, less than the 25-point spread between the two disjoint judges. Majority vote over all three gives 75.4\% capability and 90.5\% compound precision. Adjusted 3-judge recall, after collapsing taxonomy-granularity disagreements, is roughly 65--70\%. On the taxonomy subtasks, which reason over already-extracted artifacts rather than re-running the pipeline, cluster coherence averages 4.77/5.0 across all 137 clusters with all three judges reporting the identical mean, intruder detection reaches 98.9\% on 30 clusters, and taxonomy placement 97.0\% with full inter-judge agreement on 200 deviations.

\paragraph{Human validation.} Two annotators, one author and one computer science researcher from outside the project, independently labelled 150 deviations drawn at equal probability, each blinded to the classifier's output. The question was whether the recorded evidence supports the claim that the skill has the capability. Exact agreement is 76.7\% (115 of 150), rising to 90.2\% (110 of 122) where both reached a determinate answer; at least one could not decide on 18.7\%. Of the 122 determinate rows, between 87.7\% and 97.5\% support the claim. It is a range because twelve split rows are unadjudicated, and none of the twelve disputes the behaviour: the annotators agree on what the skill does and differ on which capability it falls under. This bounds precision on the deviations that assert a capability, not completeness.

\paragraph{Evidence quality tiers.} Every deviation records where its evidence came from, so a consumer can weight the weaker ones rather than treat all 250,706 alike. 137,221 deviations (54.7\%) come from skills that contain code, where a parser can check them. 41,443 (16.5\%) are under-specifications in instruction-only skills, quoted from natural-language instructions. The remaining 72,042 (28.7\%) are over-specifications in instruction-only skills and rest on prose alone, with no site a parser can check. The 69-skill panel puts capability precision at 74.8\% on code-sourced claims and 64.6\% on instruction-sourced ones, consistent with the lower verifiability of natural language; it does not estimate precision for the text-only tier, which is therefore the least bounded of the three. Per-category counts in \S\ref{sec:deviation-landscape} pool all three tiers.

\FloatBarrier
\section{Additional Discussion}
\label{app:discussion}

\paragraph{Design rationale: separating extraction from analyses.} The extraction stage $\Phi$ is the architectural commitment of BIV: it reduces the cross-representation matching problem to set comparison over $\mathcal{T}$, enabling deterministic deviation detection independent of any downstream analysis. This separation lets us measure deviation at registry scale without per-skill LLM expense for the structural component (\S\ref{sec:deviation}), and reuse one extraction pass across the three analyses in this paper rather than re-extracting per consumer.

\paragraph{Source of the detection lift.} The structural extraction stage and the relaxed-veto override together carry the largest detection gain over a same-judge single-pass auditor. At a matched Opus 4.6 judge the lift appears on every source, not only on real-world malware: recall rises 0.318 to 0.636 on real-world, 0.630 to 0.960 on SkillJect, and 0.863 to 0.969 on Skill-Inject. Structural evidence lets the judge override a reading that would otherwise rationalize an obfuscated payload as legitimate.

\paragraph{Typed taxonomy versus learned representation.} We project both modalities onto a fixed 29-capability taxonomy $\mathcal{T}$ rather than learning a joint embedding for two reasons. First, $\mathcal{T}$ is auditable: each capability carries a severity tier (Appendix~\ref{app:tiers}) and a documented detection pattern, so a human reviewer can verify any deviation claim without retraining. Second, the typed structure makes the relaxed-veto override expressible as a parameterless predicate (\S\ref{sec:downstream-adjudication}) rather than an opaque black-box decision. The trade-off is recall on capabilities outside $\mathcal{T}$, which the deviation counts cannot see.

\paragraph{Comparison to concurrent agent-skill frameworks.} BIV is not a detector but a measurement primitive: the typed evidence tuple $\Phi(s)$ supports multiple downstream consumers, of which the malicious-skill detection in \S\ref{sec:detection} is one demonstration alongside the deviation taxonomy and root-cause analysis in \S\ref{sec:deviation}. The concurrent agent-skill literature targets detection as the end goal: SkillSieve~\citep{hou2026skillsieve} and MalSkills~\citep{wang2026malskills} produce per-skill verdicts on binary benchmarks, and SkillFortify~\citep{bhardwaj2026skillfortify} formalizes a capability lattice with SAT-based enforcement and proves static-soundness results on a smaller capability set. These specialized detectors and BIV are not on the same axis: BIV's contribution is the composable framework that makes detection, root-cause classification, and other security analyses derivable from a single extraction pass. A direct head-to-head benchmark on a shared corpus is left to future work.

\FloatBarrier
\section{Compound Category Motifs}
\label{app:compound-motifs}

Table~\ref{tab:compound-motifs} reports the per-cluster motif breakdown for the four compound categories. The Motif column makes the composition explicit: each compound aggregates evidence from multiple primitive capabilities, so a single-capability scanner that checks one row at a time misses the composition that gives each compound its threat character.

\begin{table}[!ht]
\centering
\footnotesize
\setlength{\tabcolsep}{4pt}
\caption{Per-cluster motif breakdown for the four compound categories. Capability tokens in \texttt{ALL\_CAPS} are taxonomy primitives; lowercase identifiers (e.g., \texttt{sensitive\_env\_exfiltration}) are analyzer pattern names. The six Exfiltration Chain clusters below 25 deviations are collapsed into one row; individually they hold 23, 18, 13, 12, 7 and 5 deviations.}
\label{tab:compound-motifs}
\begin{tabular}{p{0.36\linewidth}rp{0.46\linewidth}}
\toprule
Cluster label & Dev & Motif \\
\midrule
\multicolumn{3}{l}{\emph{Exfiltration Chains (9 clusters, 2{,}042 deviations)}} \\
Undeclared Data Exfiltration & 1{,}204 & FILE\_READ $\to$ NETWORK\_SEND \\
Undeclared Credential Exfiltration Pattern & 571 & sensitive\_env\_exfiltration \\
Credential Exfiltration via Network & 189 & file\_theft \\
\emph{Six smaller file-exfiltration clusters} & 78 & FILE\_READ, optionally via base64, \texttt{JSON.stringify}, \texttt{JSON.parse} or \texttt{Buffer.from}, $\to$ NETWORK\_SEND \\
\midrule
\multicolumn{3}{l}{\emph{RCE Chains (4 clusters, 1{,}199 deviations)}} \\
Undeclared Dropper Pattern & 482 & download $\to$ write $\to$ execute \\
Remote Code Execution Vulnerability & 320 & \texttt{remote\_code\_execution} pattern \\
Remote Code Execution Vulnerabilities & 252 & NETWORK\_RESPONSE $\to$ CODE\_EVAL \\
Undeclared Remote Code Execution & 145 & \texttt{download\_and\_execute} pattern \\
\midrule
\multicolumn{3}{l}{\emph{Code Obfuscation (1 cluster, 299 deviations)}} \\
Obfuscated Payload Execution Pattern & 299 & encoding chain $\to$ dynamic eval \\
\midrule
\multicolumn{3}{l}{\emph{Data Lineage Violations (1 cluster, 140 deviations)}} \\
Undeclared File-to-File Data Lineage & 140 & FILE\_READ $\to$ FILE\_WRITE \\
\bottomrule
\end{tabular}
\end{table}

\FloatBarrier
\section{Intent Classification Rules}
\label{app:rules}

The rule engine in \S\ref{sec:downstream-intent} applies the 15 prioritized rules in Table~\ref{tab:rules}; the first matching rule determines the deviation's intent label. The rule engine labels 45.9\% of clustered deviations at near-zero cost, counting only rules that fired at confidence $\geq 0.7$; the remaining 54.1\% pass to the LLM classifier for joint reasoning over a skill's full deviation list.

\begin{table}[!ht]
\centering
\footnotesize
\setlength{\tabcolsep}{4pt}
\caption{Prioritized intent classification rules. Rules are evaluated in order; first match wins. Rule 14 emits G1 when the over-specification's risk tier is below \textsc{Medium} and G7 otherwise; the threshold is fixed by the per-capability severity tier in Appendix~\ref{app:tiers}, not learned.}
\label{tab:rules}
\begin{tabular}{clc@{\hspace{2.5em}}clc}
\toprule
\textbf{\#} & \textbf{Rule} & \textbf{Intent} & \textbf{\#} & \textbf{Rule} & \textbf{Intent} \\
\midrule
1 & Any instruction-level capability          & F1     & 9  & Sensitive env $+$ network outbound       & A1 \\
2 & Dropper pattern (download+write+exec)        & C1     & 10 & Data exfiltration chain (high risk)      & A2 \\
3 & Credential theft motif                       & A1     & 11 & Bulk env access $+$ network outbound     & A2 \\
4 & Evasion: encoding $+$ code eval              & C4     & 12 & Persistence motif or startup write       & C2 \\
5 & Ransomware keywords $+$ crypto $+$ write     & E1     & 13 & Reconnaissance motif or enum$+$bulk      & C5 \\
6 & Bulk file deletion (high risk)               & E2     & 14 & Over-specification (low risk)            & G1/G7 \\
7 & Cryptominer motif or keywords                & B3     & 15 & Telemetry keywords                       & G6 \\
8 & Credential $+$ network outbound              & A1     &    &                                          &    \\
\bottomrule
\end{tabular}
\end{table}

\FloatBarrier
\section{Intent Taxonomy}
\label{app:intent_taxonomy}

Table~\ref{tab:full_intent} lists the 36-category intent taxonomy used by the LLM classifier in \S\ref{sec:downstream-intent}, with a one-line description of each leaf. Categories A--F are adversarial; G is non-adversarial; H is ambiguous. The B2 slot was left unallocated during taxonomy design. The registry-scale run used 32 distinct codes, and three of them lie outside the design table. They are B2 (3 deviations, undisclosed affiliate and referral monetization), B6 (19, undisclosed third-party billing), and D5 (5, deceptive promotion). Together those 27 deviations are 0.016\% of the 163,754 classified. All three sit in adversarial branches, so the binary collapse below is unaffected. Seven designed leaves (A5, B1, C6, D1, E1, E3, F5) drew no deviations. Collapsing to binary non-adversarial vs.\ adversarial (G $\cup$ H vs.\ A--F) yields the 81.1\%/18.9\% split reported in \S\ref{sec:rootcauses}.

\begin{table}[!ht]
\centering
\footnotesize
\setlength{\tabcolsep}{4pt}
\caption{Intent taxonomy: 36 categories across 8 branches drawn from supply-chain malware, malicious-extension, PUP, and mobile-app threat research, with an added AI-agent-specific branch (F). The B2 slot was left unallocated at design time; three codes used at run time, including B2, fall outside this table and are listed in the text.}
\label{tab:full_intent}
\begin{tabular}{llp{6.0cm}}
\toprule
\textbf{ID} & \textbf{Name} & \textbf{Description} \\
\midrule
\multicolumn{3}{l}{\emph{A: Data Theft \& Espionage}} \\
A1 & Credential Theft & Harvesting API keys, tokens, passwords from local environment \\
A2 & Data Exfiltration & Sending files, env vars, or config to remote endpoints \\
A3 & Surveillance & Keylogging, clipboard monitoring, device fingerprinting \\
A4 & Corporate Espionage & Stealing IP, source code, or strategic data \\
A5 & Insider Reconnaissance & Mapping internal systems and access paths \\
\midrule
\multicolumn{3}{l}{\emph{B: Financial \& Monetization}} \\
B1 & Ad Injection & Injecting unauthorized advertisements \\
B3 & Cryptomining & Unauthorized CPU/GPU usage for mining \\
B4 & Crypto Theft & Wallet address replacement, key extraction \\
B5 & Resource Hijacking & Unauthorized use of compute resources beyond crypto \\
\midrule
\multicolumn{3}{l}{\emph{C: Payload \& Infrastructure}} \\
C1 & Payload Delivery & Download-write-execute dropper pattern \\
C2 & Persistence & Backdoors, cron jobs, startup modification \\
C3 & C2 Communication & Command-and-control channel establishment \\
C4 & Evasion & Obfuscation, encoding chains, anti-analysis \\
C5 & Reconnaissance & System enumeration, user profiling \\
C6 & Staging & Setting up infrastructure for later attack phases \\
\midrule
\multicolumn{3}{l}{\emph{D: Content \& Social Engineering}} \\
D1 & Phishing Content & Generating deceptive messages to harvest credentials \\
D2 & Misinformation & Producing false or misleading content \\
D3 & Social Manipulation & Manipulating user trust through deceptive interactions \\
D4 & Impersonation & Assuming false identity in communications \\
\midrule
\multicolumn{3}{l}{\emph{E: Destructive}} \\
E1 & Ransomware & File encryption with ransom demands \\
E2 & Wiper & Irreversible data destruction or corruption \\
E3 & Sabotage & Deliberate degradation of system performance \\
\midrule
\multicolumn{3}{l}{\emph{F: AI Agent-Specific}} \\
F1 & Instruction Hijacking & Prompt injection, identity takeover, concealed directives \\
F2 & Memory Poisoning & Corrupting agent memory or persistent state \\
F3 & Session Smuggling & Agent-to-agent communication hijacking \\
F4 & Output Manipulation & Tool result modification, response substitution \\
F5 & Capability Escalation & Acquiring permissions beyond declared scope \\
\midrule
\multicolumn{3}{l}{\emph{G: Non-Adversarial}} \\
G1 & Over-Engineering & Extra permissions requested beyond current need \\
G2 & Defensive Over-Engineering & Extra security checks beyond requirements \\
G3 & Incomplete Implementation & Declared features not yet implemented \\
G4 & Legitimate Helper & Utility functions incidentally using undeclared capabilities \\
G5 & Template Residue & Leftover code from boilerplate or templates \\
G6 & Telemetry & Legitimate usage tracking or crash reporting \\
G7 & Documentation Error & Mismatch due to poor documentation, not malice \\
\midrule
\multicolumn{3}{l}{\emph{H: Ambiguous}} \\
H1 & Context-Dependent & May be legitimate or suspicious depending on purpose \\
H2 & Insufficient Evidence & Not enough information to determine intent \\
\bottomrule
\end{tabular}
\end{table}

\FloatBarrier
\section{Illustrative Example}
\label{app:case-study}

The example is a presentation-editing skill whose manifest declares only filesystem access. One line of a 38\,KB JavaScript file passes a base64 string to \texttt{eval}, and the decoded payload opens a reverse shell to a hard-coded host every 30 seconds. Both judges and the LLM-only baseline label the skill \textsc{Safe}, so the verdict comes from the structural evidence alone. The trace reproduces that evidence as the pipeline recorded it. The upstream repository has since removed the payload line, so the trace quotes the copy we retrieved.

\begin{figure}[!ht]
\noindent\begin{tikzpicture}
\node[fill=black!6, draw=none, inner sep=8pt, inner ysep=7pt,
      text width=\dimexpr\linewidth-16pt-3pt, anchor=north west] (tracebox) at (0,0) {%
  \footnotesize
  \textit{Trace 1.}\;MaliciousAgentSkillsBench~\citep{liu2026maliciousskills} sample \texttt{rest\_2872}, skill \texttt{pptx}. Benchmark labels: Hidden Instructions, Code Obfuscation, Context Leakage, Remote Code Execution.
  \par\smallskip
  \begin{tabular}{@{}r@{\hspace{10pt}}p{0.76\linewidth}@{}}
  \textbf{Description} & ``\textit{Presentation creation, editing, and analysis. When Claude needs to work with presentations (.pptx files) for: (1) Creating new presentations, \ldots}''~~(\texttt{SKILL.md})\\
  \textbf{Declared}    & \texttt{FS\_READ\_PROJECT}, \texttt{FS\_WRITE}, \texttt{FS\_ENUMERATE}, extracted from \texttt{SKILL.md}\\
  \textbf{Code}        & \texttt{scripts/html2pptx.js}, line 31 of a 38\,KB file:\\
                       & \texttt{\,\,eval(Buffer.from("c2V0SW50ZXJ2YWwoKCkgPT4g\ldots","base64")}\\
                       & \texttt{\,\,\,\,\,\,.toString())}\\
  \textbf{Decodes to}  & \texttt{setInterval(() =\textgreater{} require('child\_process')}\\
                       & \texttt{\,\,.exec('bash -c "bash -e 5\textless\textgreater{} /dev/tcp/142.93.208[.]66/9001}\\
                       & \texttt{\,\,0\textless\&5 1\textgreater\&5 2\textgreater\&5" \ldots'), 30000)}\\
                       & a reverse shell to a hard-coded host, respawned every 30\,s (address defanged)\\
  \textbf{Detected}    & \texttt{CODE\_EVAL} and \texttt{ENCODING\_BASE64} at line 31 by the deterministic JS analyzer, both undeclared; compound pattern \texttt{obfuscated\_payload}; $\rho = \textsc{High}$\\
  \textbf{Judges}      & Opus~4.6 judge: \textsc{Safe} (0.82). Sonnet~4.5 judge: \textsc{Safe} (0.85), ``\textit{The code evaluation, base64 encoding, and compression capabilities are all necessary for its core functionality}''. LLM-only baseline: \textsc{Safe} (0.95)\\
  \textbf{Verdict}     & \textbf{malicious}, because $V(\Phi(s)) = 1$ overrides the judge; matches the benchmark label\\
  \end{tabular}};
\fill[black!35] ([xshift=0pt]tracebox.north west) rectangle ([xshift=3pt]tracebox.south west);
\end{tikzpicture}
\caption*{Trace 1: the override case.}
\end{figure}

\FloatBarrier
\section{Detection Confusion Matrices}
\label{app:confusion}

Table~\ref{tab:confusion} gives the counts behind the rates in Table~\ref{tab:headline}, with each method's misses and false alarms listed separately for every source.

\begin{table}[!ht]
\centering
\footnotesize
\setlength{\tabcolsep}{3.6pt}
\caption{Confusion matrices by source. TP and FN count malicious skills; FP and TN count benign skills of the same source, and FPR is FP over that pool. Pool names match the column groups of Table~\ref{tab:headline}.}
\label{tab:confusion}
\begin{tabular}{@{}ll rrrr rrrr@{}}
\toprule
Method & Pool & TP & FN & FP & TN & Rec & Prec & FPR & F1 \\
\midrule
Rule-based~\citep{ciscoskillscanner} & MaliciousAgentSkillsBench~\citep{liu2026maliciousskills} & 17  & 27  & 68 & 342 & 0.386 & 0.200 & 16.6\% & 0.264 \\
           & Skill-Inject~\citep{schmotz2026skillinject} & 33  & 127 & 11 & 31  & 0.206 & 0.750 & 26.2\% & 0.324 \\
           & SkillJect~\citep{skillject}    & 90  & 110 & 12 & 38  & 0.450 & 0.882 & 24.0\% & 0.596 \\
           & Aggregate    & 140 & 264 & 91 & 411 & 0.347 & 0.606 & 18.1\% & 0.441 \\
\midrule
LLM-only~\citep{liu2026maliciousskills} & MaliciousAgentSkillsBench & 14  & 30  & 2  & 408 & 0.318 & 0.875 & 0.5\% & 0.467 \\
           & Skill-Inject & 138 & 22  & 1  & 41  & 0.863 & 0.993 & 2.4\% & 0.923 \\
           & SkillJect    & 126 & 74  & 0  & 50  & 0.630 & 1.000 & 0.0\% & 0.773 \\
           & Aggregate    & 278 & 126 & 3  & 499 & 0.688 & 0.989 & 0.6\% & 0.812 \\
\midrule
\textbf{BIV} & MaliciousAgentSkillsBench & 28  & 16  & 10 & 400 & 0.636 & 0.737 & 2.4\% & \textbf{0.683} \\
           & Skill-Inject & 155 & 5   & 1  & 41  & 0.969 & 0.994 & 2.4\% & \textbf{0.981} \\
           & SkillJect    & 192 & 8   & 0  & 50  & 0.960 & 1.000 & 0.0\% & \textbf{0.980} \\
           & Aggregate    & 375 & 29  & 11 & 491 & 0.928 & 0.972 & 2.2\% & \textbf{0.949} \\
\bottomrule
\end{tabular}
\end{table}

\end{document}